\documentstyle[prl,multicol,aps]{revtex}
\input epsf.sty
\begin{document}

\title{Single parameter scaling in one-dimensional localization revisited}
\author{Lev I. Deych,$^a$ A.A. Lisyansky,$^b$ and B.L. Altshuler$^c$}
\address{$^a$Physics Department, Seton Hall University, South Orange, NJ 07052\\
$^b$Physics Department, Queens College of CUNY, Flushing, NY 11367\\
$^c$Physics Department, Princeton University and NEC Research Institute, Princeton, NJ 08540}
\maketitle

\begin{abstract}
The variance of the Lyapunov exponent is calculated exactly in the
one-dimensional Anderson model with random site energies distributed
according to the Cauchy distribution. We find a new significant scaling
parameter in the system, and derive an exact analytical criterion for single
parameter scaling which differs from the commonly used condition of phase
randomization. The results obtained are applied to the Kronig-Penney model
with the potential in the form of periodically positioned $\delta $%
-functions with random strength.
\end{abstract}


\begin{multicols}{2}

\paragraph*{Introduction}

The hypothesis of single-parameter scaling (SPS) in the context of transport
properties of disordered conductors was developed in Ref.\onlinecite{gang}.
It was suggested that there exists a single parameter, conductance $g$,
which determines a scaling trajectory $g(L)$, where $L$ is a size of a
sample. It was soon understood\cite{Anderson} that scaling in the theory of
localization must be interpreted in terms of the entire distribution
function of the conductance. In order to take the fluctuations of the conductance into account, it is convenient to consider a parameter
$\tilde{\gamma}(L)=(1/2L)$
$\ln \left( 1+1/g\right) $ instead of $g$ itself\cite{Anderson}. The main properties of this parameter are that its limit $\gamma =
\tilde{\gamma}(L\rightarrow\infty)$ is non-random \onlinecite{LGP} and that it has a normal limiting distribution for $L\gg \gamma^{-1}$\onlinecite{Mello}.   SPS in this
situation implies that the distribution function of $g$ is fully determined by the mean value of $\tilde{\gamma}(L)$. This mean value, which is the scaling parameter, is close to the limiting value $\gamma$, provided that 
$L\gg \gamma^{-1}$. 
The parameter $\gamma$ is called the Lyapunov exponent
(LE), and its inverse is the localization length, $l_{loc}=\gamma^{-1}$, of a particle's wave function. 
Below the random function $\tilde{\gamma}(L)$ will be
referred to as the finite size LE. The localization length of the Anderson model (AM) have been calculated by many authors, see Refs.\onlinecite{LGP,Wegner,Derrida}. SPS presumes that the variance, $\sigma^2$, of $\tilde{\gamma}(L)$ is not an independent parameter but is related to LE in a universal way
\begin{equation}
\sigma ^{2}=\gamma/L.  \label{SPS}
\end{equation}
This relationship was first derived in Ref.\onlinecite{Anderson} within 
the so called random phase hypothesis,
which assumes that there exists a microscopic length scale over which phases of complex transmission and reflection coefficients become completely randomized.  Under similar assumptions, Eq.~(\ref{SPS})
was rederived later by many authors for a number of different models. According to the random phase model, if the phase randomization length, $l_{ph}$, is smaller than the localization length, $l_{loc}=\gamma^{-1}$,
then Eq.~(\ref{SPS}) is valid.
An analytical derivation of the distribution function of LE for the continuous Shr\"odinger equation\cite{LGP} and numerical simulations of tight-binding AM\cite{Stone} showed that the inequality $l_{ph}\gamma \gg 1$ holds as long as disorder remains weak.  
Accordingly, SPS was believed to violate only at strong disorder, when 
$\l_{loc}$ becomes microscopic.
However, recent numerical simulations for a disordered Kronig-Penney model (KPM) \cite{Dima2} demonstrated that, contrarily to the existing picture, even when disorder is weak and $l_{loc}$ is large, a strong violation of SPS is still possible.

In this paper we demonstrate, that the condition for the validity of SPS based upon the phase randomization concept is not accurate. We study an emergence and
violation of SPS in one-dimensional systems without the assumption of 
phase randomization.  We calculate exactly the
variance of the finite size LE for AM with the Cauchy distribution of
site energies (the Lloyd model) and derive Eq.~(\ref{SPS}) without any 
{\em ad hoc} hypothesis.   This calculation also produces an analytic 
criterion
for SPS in the form $\gamma ^{-1}\gg l_{s}$. The new length scale 
$l_{s}$, which is different from the phase randomization
length, is a new significant length scale. We show that even in
the limit of weak disorder, when the localization length is
macroscopic, the states at the tails of the spectrum never obey SPS.

One-dimensional models with off-diagonal disorder
(random hopping models) represent a special case. These models demonstrate a delocalization transition in the vicinity of the zero energy state \cite
{off-diagonal}, which results in a violation of SPS\cite{off-diagonal1} as
well as in other unusual phenomena. In this paper we focus upon regular
one-dimensional situations, which include models with diagonal disorder as
well as
random hopping models far away from the critical point, $E=0$.

SPS also takes place in the regime of weak
localization, which occurs at weak disorder in two and three 
dimensional electron systems, as well as in quasi-one-dimensional wires 
in the limit of a large number of conducting channels \cite{Beenakker}. In 
this case, SPS
manifests itself in the form of universal conductance
fluctuations \cite{UCF1,UCF2}. In the weak
localization limit the mean
conductance coincides with $(\gamma L)^{-1}$ and remains the only
significant parameter, which also determines non-Gaussian corrections to 
the high moments of the distribution function  \cite{UCFreview}. SPS is
violated only for very higher moments, when the far tails of the
distribution function become important\cite{UCFreview,Muttalib}. 
This situation is also out of the scope of the present paper, since the 
weak localization limit can not be realized in one-dimensional models.

\paragraph*{Calculation of the variance of the Lyapunov exponent for an exactly solvable model}

In this paper we discuss a quantum particle on a chain of sites, which is described by the following equations
\begin{equation}
\psi _{n+1}+\psi _{n-1}-U_{n}\psi _{n}=0,  \label{model}
\end{equation}
where $\psi _{n}$ represents the wave function of the system at $n$th site.
The meaning of $U_{n}$ depends upon the interpretation of the model (\ref
{model}).  AM corresponds to Eq.~(\ref {model}) 
with $U_{n}=-E+\epsilon _{n},$ where
$E$ is the energy of the particle, and $\epsilon _{n}$ is a random site
energy.  Another realization of Eq.~(\ref {model}) is a KPM with the 
potential given by the sum, $\sum V_{n}\delta (x-na)$, of 
periodically positioned $\delta $-functions with random strengths, 
$V_{n}$.
In this case $U_{n}$ is defined as $U_{n}=2\cos (ka)+\frac{V_{n}}{k}\sin (ka)$, where $k=\sqrt{E}$ is an energy variable and $a$ is the period of the structure\cite
{LGP,Altshuler}. We assume that the parameters $\epsilon _{n}$ or $V_{n}$ are distributed
with the Cauchy probability density (the Lloyd model):
\begin{equation}
P_{C}(x)=\frac{1}{\pi }\frac{\Gamma }{(x-\xi _{0})(x-\xi _{0}^{\ast })},
\label{P}
\end{equation}
where $\xi _{0}=x_{0}+i\Gamma $, and $\xi^{\ast} _{0}=x_{0}-i\Gamma $. Parameters 
$x_{0}$ and $\Gamma $ represent the mean value and
the width of the distribution, respectively. For AM $x_{0}=0$, whereas for the KPM $x_{0}=V_{0}$.
The distribution of parameters 
$%
U_{n}$ also has the form of Eq.~(\ref{P}), where we now denote
$U_{0}$ as the center of the distribution, and $\Gamma _{U}$ as its
width. Specific expressions for $U_{0}$ and $\Gamma _{U}$ in both AM and KPM can
be obtained straightforwardly.

Below we calculate the mean value, $\gamma$, and the variance, $\sigma^{2}$,
of the finite size LE, $\gamma(L)$, in the limit 
$L\gg \gamma ^{-1}$
(further, all lengths are normalized by the lattice constant, $a$). 
Following Ref.\onlinecite{LGP} we 
present average LE as 
\[
\gamma =\left\langle \ln \left| z_{n}\right|
\right\rangle, \; \; \; 
z_{n}=\psi_{n}/\psi _{n-1} 
\]
and averaging $ \left\langle  \ldots \right\rangle $ is carried out over the stationary
distribution of the random variables $z_{n}$. It turns out \cite{LGP} that $z_{n}$ are distributed according to Eq.~(\ref{P})
with the complex parameter $\xi _{0}$ replaced by $\xi _{st}$, which 
satisfies the
equation
\begin{equation}
\xi _{st}+\xi _{st}^{-1}=U_{0}+i\Gamma _{U}.
\label{stationary}
\end{equation}
It is convenient to parameterize $\xi _{st}$ as
\begin{equation}
\xi _{st}=p\exp (i\varphi ).
\label{pfi}
\end{equation}
In this notation $\gamma =\ln \left| p\right|$. Both $p= \left|  \xi 
_{st}\right| $ and the phase, $\varphi (E)=\arg (\xi _{st})$, depend on 
the energy, $E$.

The variance $\sigma ^{2}$ can be expressed in terms of $z_{n}$ as 
\begin{equation}
\sigma ^{2}=\frac{2}{L^{2}}\sum_{n=0}^{N-1}\sum_{k=1}^{N-n}\left\langle \ln
\left| z_{n}\right| \ln \left| z_{n+k}\right| \right\rangle +\frac{1}{L}%
\left\langle \ln ^{2}\left| z_{n}\right| \right\rangle -\gamma ^{2}.
\label{variance}
\end{equation}
Therefore, $\sigma^{2}$ can be expressed through the two point 
distribution,
$P_{2}(z_{n},z_{k})$, of the parameters $z_{n}$. 
A joint distribution of multiple random
variables can be expressed in terms of the product of marginal and
conditional distributions. The latter probability distribution 
of $z_{k}$,  under the condition that $z_{n}$ is fixed, can be shown to 
satisfy the following recurrent relation ($k>n$): 
\begin{equation}
P(z_{n}|z_{k+1}) 
=\int P\left( z_{n}|z_{k}\right) P_{C}\left(
z_{k+1}+z_{k}^{-1}\right) dz_{k},  \label{Prec}
\end{equation}
where $P_{C}$ is the Cauchy distribution,
Eq.~(\ref{P}). 
The advantage of the Cauchy distribution is that the recurrence relation (\ref{Prec}) can be
solved exactly. The 
solution again has the form of the Cauchy distribution, Eq.~(\ref{P}): 
\begin{equation}
P(z_n|z_k)=\frac{\Gamma _{k-n}}{\pi }\frac{1}{\left( z_k-\xi
_{k-n}\right) \left( z_k-\xi _{k-n}^{\ast }\right) },  \label{Pcond}
\end{equation}
with complex parameters $\xi _{k}$ obeying the equation 
\begin{equation}
\xi _{k}+\xi _{k-1}^{-1}=U_{0}+i\Gamma _{U}.  \label{ksi}
\end{equation}
and $\Gamma_k = \rm{Im}( \xi_k)$. The same Eq.~(\ref{ksi}) determines parameter $\xi _{st}$ of the one-point
distribution of quantities $z_{n}$\cite{LGP}. In the latter case however, one 
looks for the stationary solution of the equation, while the
conditional distribution requires solution of Eq.~(\ref{ksi}) with the initial
condition $\xi _{0}=z_{n}$. With the use of the evaluated two-point probability
distribution, the correlator $\left\langle \ln \left| z_{n}\right| \ln
\left| z_{n+k}\right| \right\rangle $ can be represented as 
\[
\left\langle \ln \left| z_{n}\right| \ln \left| z_{n+k}\right| \right\rangle
=\frac{%
\mathop{\rm Im}
\xi _{st}}{\pi }\displaystyle\int\limits_{-\infty }^{\infty }\frac{\ln
\left| z\right| \ln \left| \xi _{k}(z)\right| }{\left( z-\xi _{st}\right)
\left( z-\xi _{st}^{\ast }\right) }dz. 
\]
One can substitute this correlator into Eq.~(\ref{variance})
and  sum it over $k$ without further assumptions.
After 
calculating the average $\left\langle \ln ^{2}\left|
z_{n}\right| \right\rangle $ 
the variance $\sigma ^{2}$
can be expressed through the parameters $p$ and $\varphi$. 
Eq.~(\ref{pfi}): 
\begin{eqnarray}
\sigma ^{2} &=&\frac{1}{L}\left\{ -\gamma \ln 
{\displaystyle{p^{4}-2p^{2}\cos (2\varphi )+1 \over \left( p^{2}-1\right) ^{2}}}%
\right.  \label{varfinal} \\
&&\left. +%
\displaystyle\int
\limits_{\varphi }^{\pi }dx\arctan \left[ \frac{\beta }{\zeta \cos \varphi
-\cos x}\right] \right\} +O\left( 1/L^{2}\right).  \nonumber
\end{eqnarray}
Here we are interested in the limit $\gamma \ll 1$, i.e., the 
localization length is large $l_{loc}\gg a=1$.
In this limit the parameters $\beta (\gamma,\varphi) $ and $\zeta 
(\gamma , \varphi) $ are equal to $\beta \simeq 2\gamma \sin \varphi $ and 
$\zeta \simeq 1+2\gamma ^{2}$, respectively.
We also assume that $\varphi \leq \pi /2$, the case $\varphi \geq \pi 
/2$ can be handled by the replacement $\varphi \rightarrow \pi -\varphi 
$. Eq.~(\ref{varfinal}) is the
main result of our calculations.  It presents the asymptotically exact ($%
L\rightarrow \infty $) expression for the variance of LE in the Lloyd 
model.
\paragraph*{Discussion}

The behavior of the integral in Eq.~(\ref{varfinal}) is governed by the
parameter 
\begin{equation}
\kappa =\left( \gamma l_{s}\right) ^{-1}, \label{kappa}
\end{equation}
where $l_{s}$ is a new length scale in the system:
\begin{equation}
l_{s}=1/\sin \varphi = |\xi_{st}|/\mathop{\rm Im}\xi_{st},  \label{ls}
\end{equation}
and  $\xi_{st}$ is given by Eq.~(\ref{stationary}). In the limit $l_s \ll \gamma$, i.e.,
\begin{equation}
\kappa \gg 1,  \label{criterion}
\end{equation}
Eq.~(\ref{varfinal}) reduces to 
\begin{equation}
\sigma ^{2}=2 \gamma /L,  \label{SPSLoyd}
\end{equation}
implying validity of SPS. Thus, Eq.~(\ref{criterion}) represents a true criterion for SPS.

It should be noted, however, that our result, Eq.~(\ref{SPSLoyd}), differs  from Eq.~(\ref{SPS}) by the factor of $2$. This difference reflects the well known peculiarity of the Cauchy distribution - all of its moments, except for the first one, diverge. As a result,
neither the approximation of the Gaussian white noise \cite{LGP}, nor a weak disorder expansion \cite{Stone} used to derive Eq.~(\ref{SPS}) can be applied to the Lloyd model. 
We have calculated the variance of LE 
for AM with the Cauchy distribution of site energies using the random phase hypothesis, and following the approach of Ref.\onlinecite{Stone}. We found 
that though a proportionality between the variance and LE persists, the numerical coefficient differs from both Eq.~(\ref{SPS}) and Eq.~(\ref{SPSLoyd}). 
This result implies that the phase randomization model is not valid at all for the Lloyd model. More important, however, is the fact that SPS  holds even when the phase randomization hypothesis fails. This is an additional confirmation of the fact that the real criterion for SPS is given by the inequality 
(\ref{criterion}) rather then by phase randomization. 

Even though the localization properties of the Lloyd model are the same as those of generic models\cite{ThoulessLloyd}, one might question the generality of the 
new scale
$l_s$ and criterion ({\ref{criterion}) in light of the 
peculiarity of the Lloyd model. 
In order to confirm a generic nature of applicability of Eq.~(\ref{criterion}), we carried out numerical simulations of KPM with rectangular barriers of
random widths.  
Statistics and the shape of potential in these calculations are considerably different from the Lloyd model. The 
results of the
simulations are shown in Fig.~1 along with $\tau (\kappa)$ obtained from our
analytical Eq.~(\ref{varfinal}). 
One can see from  Fig.~1 that the crossover
between different asymptotes occurs in the same region for both models.
This allows us to conclude that the crossover length $l_s$ and criterion ({\ref{criterion}) retain their significance beyond the Lloyd model.

According to Thouless 
\cite{Thouless}, the phase $\varphi (E)$ is proportional to the integrated
density of states $\varphi (E)=\pi G(E)$ (in the case of KPM $\varphi (E)$  must be reduced to the interval
\begin{figure}
\epsfxsize=3in \epsfbox{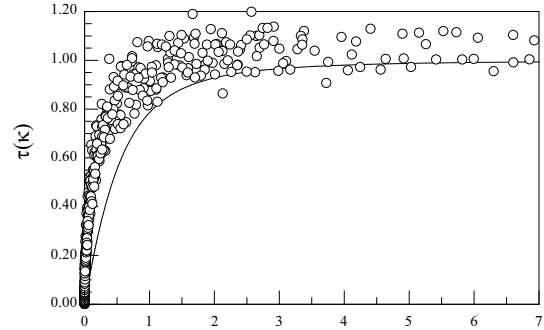}
\vspace{-2.02in}
\caption{The function $\tau(\kappa)$ obtained from the analytical solution of the Lloyd model Eq. (\ref{varfinal}) (solid line) and the numerical simulations of KPM with rectangular barriers of  random widths (open circles). Note that though the two different models give slightly different asymptotes of the  $\tau(\kappa)$ function, the crossover length is the same for both models.}
\end{figure} 
\noindent $[0, \pi]$). For the states close to the center of the initial conductivity bands, $\varphi (E)\sim \pi /2$ and consequently $l_{s}\sim 1$. Inequality (\ref
{criterion}) reduces then to $l_{loc}\gg 1$, which essentially implies that disorder is weak. 
However, as soon as the energy approaches a
band edge, $l_{s}$ grows significantly, so that
$\kappa \ll 1$ can coexist with
weak disorder, $l_{loc}\gg 1$, when $\varphi \ll 1$ or $\pi -\varphi \ll 1$.
These inequalities hold for both AM and KPM at energies outside the
initial spectrum, i.e., at the tails of the density of states for AM and within
former band-gaps for KPM. 
In this case $l_s(E)$ can be expressed in terms of the number of states between $E$ and the closest fluctuation spectrum boundary. For AM with the Cauchy distribution these boundaries are at $\pm\infty$. For KPM their role is played by energies at which $\varphi (E)/\pi$ is an integer.  The states in these
regions arise due to rare realizations of the disorder, and can be
associated with spatially localized and well-separated structural
defects. The length $l_{s}$ can then be interpreted as an average distance
between such defects.  In view of this interpretation of $l_{s}$, the transition between two types of scaling regimes occurs when the average distance between these defects becomes comparable with the localization lengths
of the respective states.

The variance  $\sigma^2$ in the limit opposite to Eq.~(\ref{criterion}) can be conveniently presented in terms of the 
parameter $\tau
=\sigma ^{2}L/(2\gamma )$: 
\begin{equation}
\tau ={\kappa }\left( {\frac{\pi} {2}}-\kappa \right)  \label{univers}
\end{equation}
It is, thus, determined by the scale $l_{s}$ rather than the localization length $l_{loc} $. This equation describes the transition from SPS to scaling with two
independent parameters $l_{loc}$ and $l_{s}$.  
The particular form of the function $\tau(\kappa) $ is probably model dependent. At the same time our results suggest that $\kappa$ is a universal parameter, which naturally describes the crossover between different types of statistics in the localization problem.

The previous criterion for SPS based upon the phase randomization length\cite
{Anderson,Stone} effectively restricts the strength of disorder\cite{LGP,Stone}. The
new criterion, Eq.~(\ref{criterion}), separates all the states of the system with a given strength of 
the disorder into two groups, which
 demonstrate distinct scaling properties. The boundary between the groups, defined by the condition $\kappa =1$, coincides
with the boundaries of the initial spectrum $\left| U_{0}\right| =2$, provided
that $\Gamma _{U}\ll 1$. For the states inside the original spectra, SPS holds as long as the disorder 
remains weak. The
length scale $l_{s}$ in this case is of the order of the lattice constant
$a$ and, thus, has no scaling significance. Nevertheless, it is important to notice that this length scale differs from the phase randomization length. Indeed, it was found numerically in Ref.~\onlinecite{Stone} that  $l_{ph}$ increases toward the
center of the band $E=0$ as $1/E$. Absence of the phase randomization for $E=0$ was also obtained analytically in Ref.\onlinecite{Dmitriev}.  Both numerical,  Ref.\onlinecite{Stone}, and analytical, Ref.\onlinecite{Dmitriev}, calculations suggest that SPS still holds at $E=0$, 
in spite of the fact that the phase does not randomize and the standard 
weak-disorder
expansion for LE fails \cite{Wegner,Derrida} at this energy.
At the same time, applying criterion (\ref{criterion}) one finds that SPS 
in this case violates only when disorder is so strong that the 
localization length becomes of a microscopic magnitude, $l_{loc}\simeq 1$ 
in accord with Ref.\onlinecite{Stone}.
Indeed, $l_{s}\simeq 2/\sqrt{4-U_{0}^{2}}$ and decreases toward the 
center of the band, where it assumes the minimum value $l_{s}=1$.  This allows us to conclude that initially conducting
states always demonstrate SPS in a meaningful scaling regime $l_{loc}\gg 1$.

The situation for the states from the former band-gaps is totally
different: SPS violates when both lengths, $l_{s}$ and 
$l_{loc}$, are macroscopic. Such a violation is significant from the 
scaling point of view.
In the case of KPM $l_{loc}$ remains macroscopic throughout entire
band-gaps, provided that the energy is sufficiently high, $k\gg V_{0}$. 
Contrarily,  $l_{loc}\sim 1$ for those states of  AM, that are
not very far from the boundary. A strong violation of SPS in this case
coincides with the system being driven out of the scaling regime.

Change in the strength of disorder affects the two scaling lengths
$l_{loc}$ and $l_{s}$ differently. In the case of the gap states, 
$l_{loc}$
only weakly depends upon disorder. It is approximately equal to the 
penetration length,
which would describe tunneling through the system in the absence of 
disorder. Disorder related corrections to this quantity are of the order 
of $\Gamma _{U}^{2}$. The parameter $l_{s}^{-1},$ on the contrary, is 
linearly proportional to $\Gamma _{U}$. Increase in disorder, therefore, 
causes the critical parameter $\kappa $ to increase. Thus, the 
system can crossover between $\kappa \ll 1$ and SPS ($\kappa \gg 1$) 
regimes with an enhancement of the disorder. The results seem paradoxical 
since the restoration occurs because of increasing disorder, which must, 
however, remain small enough to keep the system within the scaling 
regime. This effect is also more important for KPM than for AM, and was 
observed numerically in Ref.\onlinecite
{Dima2} for a periodic-on-average random system.

Another interesting feature, which is specific for KPM and absent in AM, 
is a presence of resonance states where $\gamma =0$ regardless of the 
disorder. Though this feature is
unstable with regard to a violation of the strict periodicity in the
position of the $\delta -$functions, it is also present in some other 
models,
such as random superlattices, and deserves a discussion. At the 
resonance, both 
$\gamma $ and  $l_{s}^{-1}$ vanish. Their
ratio, $\kappa $, however, remains finite and experiences a jump at the
resonance point  from the value of $2V_{0}/\Gamma \gg 1$ at the band side 
of
the resonance to $\Gamma /2V_{0}\ll 1$ at the gap side. It means, that 
the change
of statistics of LE at the resonance states from SPS behavior at the band
side to two parameter scaling at the gap side occurs discontinuously.

We are indebted to A. Mirlin for a useful discussion. We also wish to 
thank
S. Schwarz for reading and commenting on the manuscript.  Work at Queens
College was supported by a CUNY collaborative grant and PSC-CUNY research
award, work at Princeton University was supported by ARO under contract 
DAAG
55-98-1-0270.

\newpage
}
}
\end{multicols}
\end{document}